# Torque and temperature dependence of the hysteretic voltage-induced torsional strain in tantalum trisulfide

H. Zhang, J. Nichols, and J.W. Brill

Department of Physics and Astronomy, University of Kentucky,
Lexington, KY 40506-0055

**Abstract:** We have measured the dependence of the hysteretic voltage-induced torsional strain (VITS) in crystals of orthorhombic tantalum trisulfide on temperature and applied torque. In particular, applying square-wave voltages above the charge-density-wave (CDW) threshold, so as to abruptly switch the strain across its hysteresis loop, we have found that the time constant for the VITS to switch (at different temperatures and voltages) varied as the CDW current. Application of torque to the crystal could also change the VITS time constant, magnitude, and sign, suggesting that, at least in part, the VITS is a consequence of residual torsional strain in the sample which twists the CDW. Application of voltage changes the pitch of these CDW twists, which then act back on the lattice. However, it remains difficult to understand the sluggishness of the response.

## I. INTRODUCTION

Quasi-one-dimensional conductors with sliding charge-density-waves (CDWs) are best known for their many unusual electronic properties, associated with polarization and motion of the CDW for applied voltages V ≥ $V_T$, the CDW depinning threshold voltage.[1,2] CDW depinning can also affect the crystal's mechanical properties, e.g. lattice strains[3,4] and drops in some elastic moduli.[5-9] For example, in orthorhombic tantalum trisulfide (o-TaS$_3$),[10] the best studied material, the Young's modulus and shear modulus decrease by ~ 2 % and ~ 25%, respectively,[5,6] and there are hysteretic changes in sample length ($\Delta L/L$ ~ $10^{-6}$).[3] The elastic anomalies have been understood as resulting from changing strain in the crystal causing relaxational changes of CDW phase domains,[7,9] while the length changes are associated with CDW polarization (i.e. rarefaction and compression on the two sides of the crystal)[11] coupling to and straining the lattice.[3,12] For an o-TaS$_3$ crystal a few mm long at temperature T ~ 80 K, the elastic relaxation time is ~ 1 sec near threshold[9] while the relaxation time for longitudinal deformations of the CDW, measured electro-optically, is at least two orders of magnitude shorter.[13]

In 2007, Pokrovskii et al reported that crystals of o-TaS$_3$ also exhibit small ($\Delta\phi$ ~ 1°) hysteretic twists when the CDW is depinned, with voltage dependences similar to that of the length changes and the CDW compressions/rarefactions.[4] (Examples of the hysteresis loops are shown in Figure 4, below.) While these hysteretic twists are very sluggish, as discussed below, they also observed much smaller and faster, reversible twists, which grew continuously with voltage with no change at threshold.[14] Similar effects were observed in other CDW conductors.[14] Since these materials and CDWs have no known polar axes, there was no clear explanation for this unique "voltage-induced torsional strain" (VITS). (We note that recently chiral structure, associated with three equivalent CDW wave vectors, has been observed in the CDW in TiSe$_2$,[15] but the CDW remains pinned at high electric fields in TiSe$_2$, so the various anomalous electronic and electromechanical properties associated with depinning have not been studied. In contrast, o-TaS$_3$ has a single CDW wave vector[16] and no known chirality.)

In our earlier work, we verified the hysteretic VITS effect in o-TaS$_3$ and studied its voltage, frequency, and time dependence (at T = 78 K).[17,18] We found that near threshold, the time constant for the VITS to jump across the hysteresis loop (i.e. when switching the applied voltage between $+V_T$ and $-V_T$) is ~ 1 s, but it decreases rapidly with increasing voltage.[18] Complete evolution of the hysteresis loops was even slower, as would be expected for voltage dependent relaxation times and strengths. We suggested that the hysteretic VITS effect was due to CDW wave fronts being twisted, even without applied voltage, e.g. due to contacts or defects.[18]

To test this hypothesis, we sought to twist the sample with an additional applied torque. Our measurements are done by placing the sample in an RF cavity[19] with a small magnetized steel wire glued to its center. When the sample twists, it moves the wire, modulating the resonant frequency of the cavity. By applying a dc magnetic field, we could also add an external torque to the sample. In addition, by simultaneously applying a small ac magnetic field, we could use the cavity response to (roughly) normalize our VITS signal. (In References 17 and 18, the VITS signals were all presented as relative values.) Our techniques are discussed in detail in Section II.

In carrying out these measurements, we observed that external torque could have a strong effect not only on the magnitude of the hysteretic VITS, as anticipated, but also on its time constant, Most surprisingly, the external torque was even observed to reverse the sign of the voltage-induced torsional strain. These results and their implications are discussed in Section III. Finally, in Section IV we discuss temperature dependent measurements carried out to try to determine the origin of the very long hysteretic VITS time constants.

## II.  EXPERIMENTAL TECHNIQUES

The techniques we used to study the hysteretic VITS effect were similar to those used in Reference 18. Electrical contacts were glued, with silver paint, to the ends of an *o*-TaS$_3$ crystal, with typical dimensions ~ 4 mm x 10 μm x 2 μm. A thin gold film was evaporated along half the length of the sample, electrically shorting this half of the sample (see Figure 1 in Ref. 20) and keeping the CDW pinned there, while the CDW could be depinned by applied voltage on the other half.[1,10] A magnetized steel wire (1-3 mm) was glued to the center of the sample (at the edge of the gold film). The sample was placed in a helical resonator RF cavity (with resonant frequency ~ 430 MHz and Q ~ 300),[19] with the end of the magnetic wire about ~ ¼ mm from the tip of the helix, as shown in the lower inset to Figure 1. When the sample twisted, it changed the helix-wire separation and hence the resonant frequency of the cavity. When driving the cavity at or near resonance, the transmitted signal would be modulated by the motion of the wire. The cavity was placed in a Helmholtz coil, so that a small magnetic field, parallel to the helix tip, could be applied by coil current $I_B$ ($B/I_B$ = 80 Gauss/Amp).

Three different types of experiments were performed: i) An ac-magnetic field was applied so that the sample would oscillate, with amplitude proportional to its shear compliance (J), allowing the voltage dependence of the shear compliance to be measured.[8,9] The cavity was driven at resonance, so that the oscillating sample phase-modulated the output of the cavity at the magnetic field frequency, giving an ac signal ($V_J$) which was measured with a lock-in amplifier.[19] As mentioned above, the compliance increases (by over 20% at low frequencies) for $|V| > V_T$, so the threshold field was determined by this experiment.[8,9] ii) A symmetric square wave voltage at frequency ω was applied to the sample, twisting the sample through the VITS effect, and the phase- modulated response of the cavity at ω measured as a function of square-wave voltage and frequency.[17,18] As in Ref. (18), we denote this complex, frequency dependent torsional strain $\varepsilon_\omega$. iii) A symmetric-triangle wave voltage was applied to the sample, sweeping the sample through a hysteresis loop. To measure the time-dependent VITS signal, the cavity was driven slightly off-resonance with an FM signal,[8] and the response at the FM frequency measured and averaged with a digital oscilloscope.[18] (Applying gold to half the sample effectively puts a voltage independent spring in parallel with the uncoated half of the sample, roughly halving the measured elastic and VITS anomalies.)

For the experiments discussed in Section III, the steel wire was ~ 3 mm long, a few times longer than that used in References 17 and 18, decreasing the torsional resonant frequency of the sample to ~ 100 Hz, but allowing us to twist the sample several degrees by an applied dc magnetic field. Since the response of the cavity to sample motion should be (approximately) inversely proportional to the helix/wire separation, we expect $1/V_{J0}$ to vary linearly with magnet current, where $V_{J0}$ is the "pinned" (i.e. V=0) compliance signal. (Because sample strains can become frozen in the sample for $|V| < V_T$, it is necessary to first depin the sample, applying $V>V_T$ at each magnetic field, before measuring $V_{J0}$.[8]) Typical results are shown in the upper inset to Figure 1, where $1/V_{J0}$ is plotted as a function of magnet current. The hysteresis shows that the sample tended to stick slightly and undershoot its "equilibrium" position. Using the measured length of the wire and (room temperature, $I_B = 0$) helix/wire separation, the field dependence of the twist angle could be determined: for sample E, $\partial\varphi/\partial I_B \sim 12°$/amp. Then, comparing the square-wave signals (experiment ii) at each magnetic field with $V_{J0}$, the voltage-induced twist angles ($\varepsilon_\omega$) could be calculated. Finally, these magnetic field dependent values of $\varepsilon_\omega$ could be used to normalize the FM signals of experiment iii. Note that all these normalizations are only approximate (~ factor of 2), in view of estimates in the sample geometry and the assumption that the helix-wire separation does not change significantly with temperature.

In Section III, we show the magnetic field dependence of the compliance, square-wave response, and hysteresis loops for two samples at T = 78 K. The general features discussed for these samples were observed for a few other samples. However, for most samples studied, the VITS responses were more complicated functions of voltage and/or frequency than for these, in some cases changing sign with increasing voltage. Possible reasons for such complex behavior include a) the presence of more than one threshold voltage, e.g. due to imperfect screening by the gold film, b) complicated residual twists in the sample, as discussed below, and c) larger than usual reversible, non-hysteretic voltage-induced twists.[14] As mentioned above, the latter grow continuously with voltage, with no threshold behavior, and could overwhelm the hysteretic VITS signal, especially for samples with large threshold voltages. These samples were rejected, as the *hysteretic* VITS effect is the subject of our study.

In Section IV, we discuss the temperature dependence of $\varepsilon_\omega$. Since the hysteretic response gets faster at higher temperature, a short (~ 1 mm) wire was attached to the sample to keep its resonant frequency high (730 Hz). Therefore, the magnetic field response of this sample was weak and, although $\varepsilon_\omega$ could still be normalized to $V_{J0}$ at each temperature, the corresponding twist angles were not calculated.

To avoid confusion with the four samples discussed in References 17 and 18, the samples discussed in this paper are named E, F, and G.

## III. APPLIED TORQUE DEPENDENCE

Figure 1 shows the dc voltage dependence of the resistance and change in shear compliance, with a 10 Hz oscillating magnetic field, for Sample E at T = 78 K, for two different dc magnetic fields which twist the sample. Note the following:

a) The resistance is independent of magnetic field.

b) There is no clear sign of the threshold voltage in the resistance data, as the resistance appears to change continuously with voltage at all voltages. This is a common problem for $o$-TaS$_3$ at low temperatures, where CDW creep commences at a second threshold below $V_T$.[2,10,21] Identifying the threshold from the resistance curve is further complicated in a two-probe measurement, because CDW phase-slip affects the I-V curve.[2]

c) The threshold field is clearly observed in the shear compliance data as the voltage at which J starts increasing.[8,9] $V_T \sim 180$ mV is independent of magnetic field within our sensitivity.

d) The change in compliance with voltage *appears* to be slightly magnetic field dependent. It is not yet clear if this is a real effect (for example, longitudinal strains are known to affect the change in shear compliance[22]) or a consequence of a nonlinear dependence of the measurement sensitivity on changes in sample position. However, if the latter, the small changes in sensitivity (~ 1%) will not have a significant effect on the square-wave results, as the relative scatter in $\varepsilon_\omega$ is > 1%.

Figure 2a shows the dependence of $\varepsilon_\omega$ on square-wave amplitude at two different magnetic fields (for $\omega/2\pi = 10$ Hz); the response both in-phase with the square-wave (solid symbols) and in quadrature (open symbols) is shown. The magnitude of the VITS is much smaller for $I_B = -0.6$ A than for $I_B = +0.3$ A, but the most striking feature is that $\varepsilon_\omega$ has opposite signs at the two magnetic fields. In addition, the peak in the quadrature signal, for which the average relaxation time (defined below) $\tau_0 \sim 1/\omega$, occurs closer to threshold for $I_B = -0.6$ A than for $I_B = +0.3$ A, implying that, at each voltage, the response is faster for $I_B = -0.6$ A.

Figure 2b shows the frequency dependence of $\varepsilon_\omega$ for two square wave voltages and magnet currents (for which $\varepsilon_\omega$ is positive). The curves show fits to the modified relaxation expression[23]

$$\varepsilon_\omega = \varepsilon_{\omega 0}/[1+ (-i\omega\tau_0)^\gamma], \quad (1),$$

where $\tau_0$ is the average relaxation time and a value of the exponent $\gamma < 1$ allows for a distribution in relaxation times; the distribution of relaxation times is given by[18,23]

$$a(\tau) = (\varepsilon_{\omega 0}/\pi)\,(\tau/\tau_0)^\gamma \sin(\gamma\pi) / [1 + 2(\tau/\tau_0)^\gamma \cos(\gamma\pi) + (\tau/\tau_0)^{2\gamma}]. \quad (2).$$

The magnetic field dependence of the fitting parameters for these two square-wave voltages is shown in Figure 3. For both square-wave voltages, the relaxation strength and average relaxation time have strong dependences on magnetic field, but whereas the relaxation strength falls monotonically with $I_B$, the dependence of $\tau_0$ differs at the two voltages. We will discuss a consequence of these dependences later. (In these fits, the exponent $\gamma$ varies from 1 to 0.65, which corresponds to a distribution of relaxation times over a decade wide.)

Figure 4 shows the magnetic field dependence of hysteresis loops for sample F, for which $\partial\varphi/\partial I_B \sim 5°$/amp. All these loops were measured with 0.3 Hz, 0.75 V triangle waves, slow enough that the shapes/sizes of the loops are close to their static limits.[18] Note that, as discussed in Reference 18, the loops are not symmetric functions of voltage; for this sample, the loop closes more gradually at positive voltages than at negative. The magnitude of the VITS, i.e. the height of the hysteresis loop, is again a strong function of torque on the sample. As $I_B$ increases toward 0.8 A, the main loop closes, leaving a subsidiary loop at negative voltage. For $I_B > 0.8$ A, the main loop starts opening again, but has now reversed direction; this corresponds to the change in sign of $\varepsilon_\omega$ for sample E shown in Figure 2a.

Note that one expects these hysteresis loops to change shape with changing applied torque due to the (~ symmetric in voltage) increases in shear compliance for $|V| > V_T$.[6,8,9] Since the changes in J are not hysteretic, they would simply add a "∩" shape to the hysteresis loops, with amplitude increasing with increasing $I_B$. For sample F, J changed by ~ 3% at V = 0.75 V, so a "∩" with amplitude comparable to the width of the largest hysteresis loop is expected at $I_B$ = 1 A. Comparison of the shapes of the loops show that this is generally seen, although the curvature does not vary regularly with $I_B$ (e.g. the "∩" curvature is a maximum at $I_B$ ~ 0.7 A), perhaps because of $I_B$ dependent values of ΔJ(V).

That the sign of the hysteretic voltage-induced torsional strain, as well as its magnitude, depends on the magnetic field, and therefore applied torque and twisting of the sample, suggests that residual twisting of the sample, even with no applied torque, is responsible for the hysteretic VITS. This residual twisting may be a consequence of how the sample is mounted on the contacts and how the magnetic wire is attached to the sample, but it may also be "built in" to the crystal; in particular the thin $o$-TaS$_3$ crystals are notorious for their large number of defects which have prevented determination of the crystal structure.[16] If a crystal has a local twist β = ∂φ/∂z, then, to first order, the local CDW wave vector will have an azimuthal component:

$$\mathbf{q} = q_0 (\underline{\mathbf{z}} + \beta r \underline{\boldsymbol{\varphi}}), \qquad (3)$$

where $q_0$ is the local wave vector in the absence of twisting, $\underline{\mathbf{z}}$ and $\underline{\boldsymbol{\varphi}}$ are unit vectors in the longitudinal and azimuthal directions, r is the radial distance from the center of the sample, and we have assumed a circular cross-section for simplicity. With application of voltage, the CDW will become polarized, becoming compressed and rarefied on the two ends of the sample,[11] changing the helical pitch:

$$q_0(z) = q_{00} + \Delta q_0(z). \quad (4)$$

$\Delta q_0(z)$ consists of both reversible, small changes close to the contacts and a long-range hysteretic component.[24,25] The latter is frozen in the sample if the voltage is removed and reverses sign when a voltage near threshold of opposite polarity is applied; i.e. it exhibits hysteresis similar to that of the voltage-induced torsional strain. As mentioned above, this component of $\Delta q_0(z)$ can cause local, hysteretic longitudinal stresses in the crystal. We similarly assume that the hysteretic changes in the azimuthal component can put torsional stress on the sample and cause the VITS.

However, the net changes in length caused by $\Delta q_0(z)$ are very small ($\Delta L/L \sim 10^{-6}$)[3] because, while the CDW deformations in *o*-TaS$_3$ have been observed to be slightly larger on the negative side of the crystal than the positive,[26] the asymmetry is small [i.e. $\Delta q_0(z) \sim -\Delta q_0(L-z)$, where L is the length between contacts of the crystal], so the compressions and stretches on the two sides of the sample almost cancel.[3,12] If the torsional stress was simply proportional to $\beta \Delta q_0(z)$, then (for constant $\beta$) the VITS would change sign in the center of the sample, with a net $\Delta \phi(L) \sim 0$ at the free end. On the other hand, if the torsional stress was proportional to $\partial q/\partial z$, the VITS would grow continuously with the distance from the clamped end, as observed.[4]

One way to accomplish this dependence on $\partial q/\partial z$ is to assume that the torsional stress that results from $\beta \Delta q_0(z)$ acts as a local external torque, $\eta$, which is opposed by the torsional rigidity $\kappa \sim GR^4/z$, where $G = 1/J$ is the shear modulus,[6] R is the effective radius of the sample, and we explicitly assume that the sample is clamped at z=0. From Eqtns. (3) and (4),

$$\eta(z) \sim (\mu/q_{00}) \int dA\, r\, (\beta r\, \Delta q_0) \sim (\mu R^4 \beta/q_{00})\, \Delta q_0 \quad (5),$$

where A is the cross-sectional area and $\mu$ is the torsional "trans-modulus" relating crystal stress to CDW strain. The change in twist angle along the length of the sample will be:

$$\partial \Delta\varphi/\partial z \sim \partial(\eta/\kappa)/\partial z \sim \mu\, \beta\, (z\, \partial q/\partial z + \Delta q_0)/\, Gq_{00}. \quad (6)$$

Consider the case of a sample with a uniform residual twist, $\beta$ = constant, and taking $\Delta q_0(z) \sim -\Delta q_0(L-z)$, the integral of the second term in (6) will approximately vanish and the twist angle of the wire at the "free" end of the sample will be given by:

$$\Delta\varphi(L) \sim \mu\, \beta\, L\, \Delta q_0(L)/Gq_{00}. \quad (7)$$

For example, the hysteresis loop of Sample F closes at $I_B = 0.8$ A, so we take $\beta L \sim 4^\circ$. Taking $G \sim 5$ GPa,[6] $\Delta\phi(L) \sim 0.1^\circ$, and $\mu \sim 40$ GPa, the value of the *longitudinal* trans-modulus found in Reference 12 (where it is called $gY_c$), we find $\Delta q_0(L)/q_{00} \sim 3 \times 10^{-3}$. This is the same relative shift in q found from transport measurements in $NbSe_3$.[11]

Of course, all these values should only be considered order of magnitude estimates. Most samples presumably have non-uniform residual twists (i.e. spatially dependent values of β), which can give rise to the complicated voltage dependences of the VITS as $\Delta q_0(z)$ varies with voltage, observed for some samples. It should also be noted that our result, in which residual twist replaces the need for a fixed polar axis in the crystal, seems to contradict one experiment done in Reference (4), in which when a sample was cut and flipped over, its VITS direction also reversed. (Note that if β is caused by growth defects rather than sample mounting, it does not change sign when the sample is flipped over.) However, given the flexibility of the crystals, one cannot rule out that cutting and remounting the sample in these experiments may have changed the sign of β and the resulting VITS. Alternatively, it is also possible that the β-dependent VITS only represents one possible mechanism, and that $o$-$TaS_3$ crystals do contain a polar axis (e.g. because of an undetected chirality[15] or surface pinning of the CDW[4,18]) that also contribute.

Our model has interesting implications if a sample were mounted so that it was free to turn at both ends. For an applied voltage above threshold, the two ends would turn in opposite directions, until stopped by its torsional rigidity, but for a uniform change in q caused by a change in temperature, both ends would turn in the same direction, with no internal restoring force. Of course, the long time constants associated with the VITS indicate that there are large internal frictional forces, not addressed by our model, which will damp the motion.

Indeed, it is difficult to understand the long time constants associated with the VITS. Near threshold, the time constant for longitudinal changes in q (i.e. CDW polarization) is governed by diffusion, with a diffusion constant inversely proportional to the square of the phason velocity;[27] at higher voltages, CDW phase-slip allows the local wave vector to change more quickly.[11,28] A sample a few times shorter than those studied here was observed to have a polarization time constant ~ 1 ms near threshold.[13] Even correcting for the $L^2$ dependence of the diffusion time, we expect the CDW polarization in our samples to change two orders of magnitude faster than does the observed VITS.

## IV. TEMPERATURE DEPENDENCE

To try to shed light on the slow torsional response, we studied the temperature dependence of the square-wave response of Sample G. As mentioned above, this sample had a shorter magnetic wire to give it a higher resonant frequency (730 Hz) so that its dynamics could be studied over a wider range (0.1 Hz $\leq \omega/2\pi \leq$ 200 Hz). Therefore, measurements of its dc magnetic field dependence were not done, although ac magnetic fields could still be used to study the voltage dependence of its shear compliance.

Figure 5a shows the dc voltage dependence of its resistance and shear compliance (with 10 Hz oscillating magnetic field) at temperatures between 90 K and 120 K. At T = 78 K (not shown), the voltage dependence of its resistance shows no threshold dependence, as discussed above for sample E. However, at higher temperature, the voltage at which the resistance falls due to CDW current is clearer, although the "resistance threshold" typically seems slightly greater than $V_T$, the threshold observed for the compliance, as discussed in Reference 9.

Figure 5b shows the 10 Hz square-wave response at the same temperatures. Note that, at each temperature, the onset voltage for the square-wave response ($V_{on}$) is slightly below $V_T$, as discussed in Reference 17. The temperature dependences of $V_T$ and $V_{on}$ are plotted below in Figure 7b; $V_T$ and $V_{on}$ are weakly temperature dependent between 90 K and 120 K, and their difference is small (25 ± 5 mV), but $V_T$ grows rapidly at lower temperatures.

To compare the dynamic response at each temperature, one should choose appropriate voltage criteria, e.g. so that there would be a fixed driving potential on the CDW. In particular, it wasn't clear whether we should use $V_T$ or $V_{on}$ as the relevant "threshold" (although since $V_T$-$V_{on}$ is approximately constant for T ≥ 90 K, the distinction isn't very important here). We therefore took measurements at the following square-wave voltages: $V_T$, $V_{on}$ + 50 mV, $V_T$ + 50 mV, $V_{on}$+100 mV, and $V_T$ +100 mV at several temperatures between 78 K and 120 K; at higher temperatures, the response moves out of our frequency window. Two examples, at $V_{square}$ = $V_{on}$+100 mV = 170 mV, are shown in Figure 6, with fits to Eqtn. 1. ($V_{on}$ = 70 mV at both 90 K and 110 K.) The increase in the speed of the VITS with increasing temperature is evident, as the peak in the quadrature response of $\varepsilon_\omega$ increases from ~ 1 Hz at 90 K to ~ 50 Hz at 110 K.

The parameters of the fits for all five voltages and temperatures 90 K < T < 120 K are plotted in Figure 7a. (At T = 78 K, the average relaxation times, even for $V_T$ + 100 mV, were so slow that we could not do meaningful fits for data in our frequency window.) For each voltage criterion, the magnitude of the VITS does not vary much with temperature in this range, consistent with the results of Pokrovskii et al.[14] The quadrature peaks broaden considerably at the lowest voltages, so the values of the exponents decrease from ~ 0.7 (corresponding to a one decade width in the time constant distribution), to 0.3 (corresponding to almost a five decade width).

For each voltage criterion, the average relaxation time falls by two decades between 90 K and 120 K. In contrast, the low-field (i.e. pinned CDW) resistance, $R_0$ only falls by a factor of ~ 3. The current carried by the CDW ($I_{CDW} = I_{total} - V/R_0$), however, increases by two decades for each voltage above $V_T$, as shown in Figure 7c, where we also plot the temperature dependence of $I_{CDW} \tau_0$. Within the ranges measured, $I_{CDW} \tau_0$ is roughly *independent of both temperature and voltage*, suggesting that the time constant of the VITS is determined primarily by the CDW current. (The temperature dependence of the relaxation time for longitudinal CDW deformations has not been measured for $o$-TaS$_3$, but for quasi-one dimensional $K_{0.3}MoO_3$, "blue bronze", it has been observed to have a much weaker dependence on both temperature and CDW current[25] than that we are observing for the VITS in $o$-TaS$_3$.)

As mentioned, at T = 78 K the VITS time constants[18] are much longer than the time constants associated with longitudinal CDW deformations near threshold;[13] comparison of the results in References 13 and 18 shows that this remains true for voltages at least up to $3V_T$. Our present results therefore suggest that, as the CDW deforms under applied voltage, sample strain is held back until "released" by the flow of CDW current. This, in turn, suggests that it is not crystalline defects (e.g. dislocation lines) that are hindering the motion, as they are not expected to interact directly with CDW current, but CDW defects, e.g. local phase deformations,[24] which are responsible. Note that for a twisted sample, there will presumably be azimuthal CDW current, parallel to the local CDW wave vector given by Eqtn. (1). If it was this azimuthal CDW current that released the strain, then one would expect that, as the sample was twisted by the applied magnetic field, the VITS relaxation time would vary inversely with its magnitude. However, as shown in Figure 3, this is not so; e.g. at $V_{square}$ = 400 mV, both $\tau_0$ and $\varepsilon_{\omega 0}$ decrease with increasing $I_B$. (Supporting the fact that azimuthal currents are not relevant is also the fact that longitudinal sample strains, as measured by the length of the crystal, also responded very sluggishly, with time constants > 1 sec, to changes in the polarity of applied voltage.[3]) Additional experiments on the temperature and current dependence of both the VITS and CDW deformations, especially transverse deformations, would be desirable to clarify their relationship, including measurements that compared their onset voltages, e.g. what limits the VITS dynamic response at voltages below the CDW current threshold. Unfortunately, such detailed measurements of local CDW deformations (e.g. using electro-optic techniques[13]) would require samples a few times wider than those that have been grown to date.

In conclusion, we have found that twisting the sample by an applied torque can affect both the magnitude and sign of the voltage-induced torsional strain, and have suggested a model in which the hysteretic VITS is due to twists in the sample causing azimuthal deformations of the CDW, which in turn change under applied voltage and then feed back on the crystal, changing its torsional strain. It is difficult, however, to account for the sluggishness of the VITS signal (e.g. at least two orders of magnitude slower than changes in CDW deformations at T = 78 K). While our measurements on the temperature dependence of the VITS suggest that it is controlled by CDW current, the mechanism for this is unclear.

We thank R.E. Thorne of Cornell University for providing samples. This research was supported by the National Science Foundation under Grants Nos. DMR-0800367 and EPS-0814194.

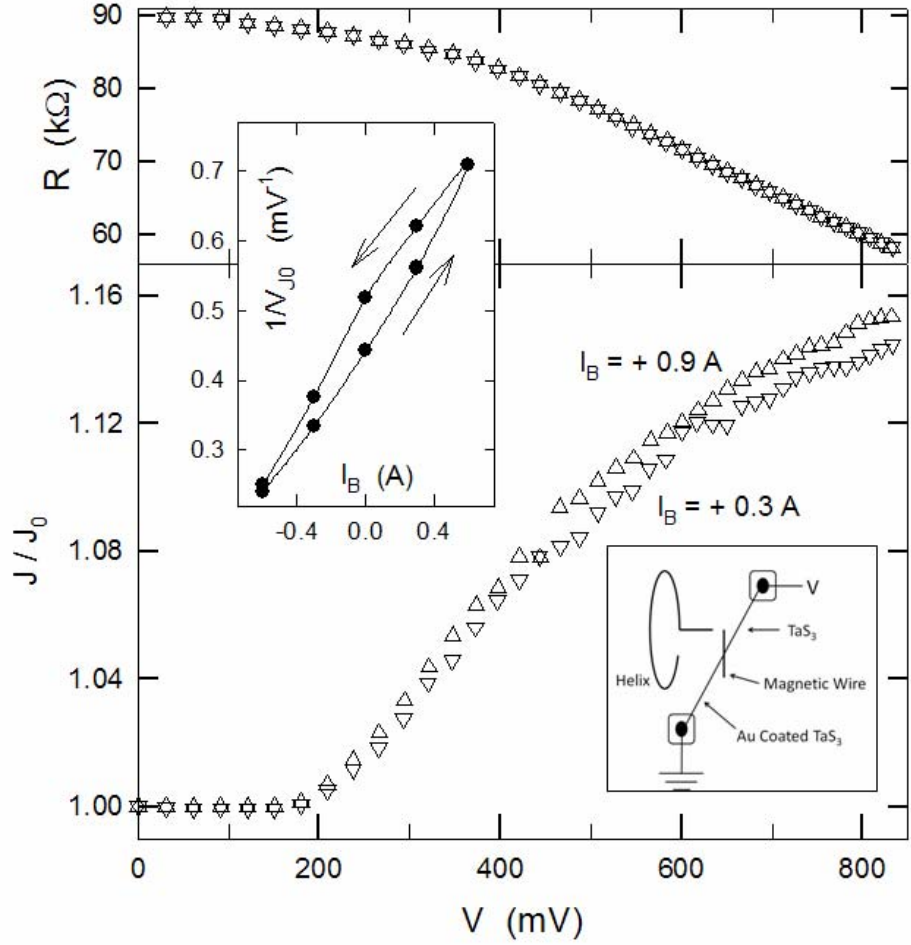

FIG. 1. Resistance (R) and shear compliance (J) vs. dc voltage across sample E at T = 78 K at two different magnetic fields. The compliance was measured with a 10 Hz oscillating torque. (Note that the symbols completely overlap for the resistance.) Upper inset: Reciprocal of $V_{J0}$, the shear compliance signal at V=0, vs. magnet current, used to find the resulting twist, $\partial\varphi/\partial I_B \sim 12°$/amp . Lower inset: schematic of the sample configuration.

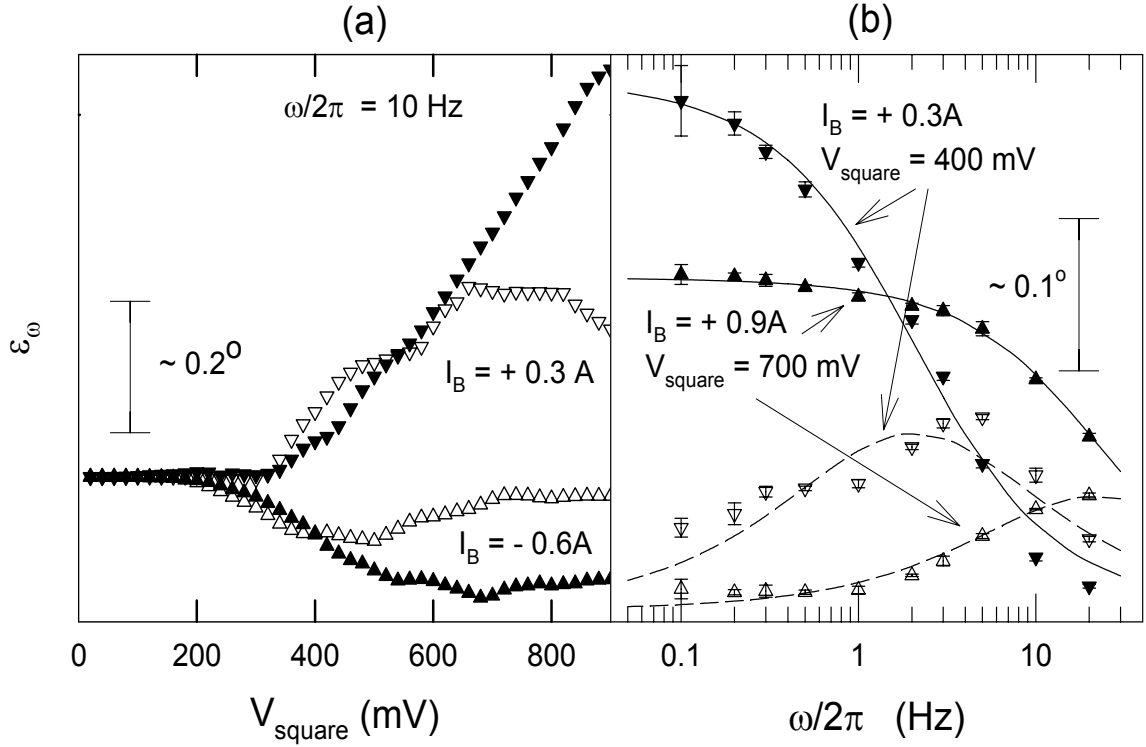

FIG.2. (a) Dependence of the VITS, $\varepsilon_\omega$, on square-wave amplitude at two magnet currents for Sample E at T = 78 K (with twist angle $\partial\varphi/\partial I_B \sim 12°/\mathrm{amp}$). Solid symbols: response in-phase with the 10 Hz square-waves; open symbols: response in quadrature with the square-waves. b) Square-wave frequency dependence of $\varepsilon_\omega$ as function of frequency for two different square-wave amplitudes and magnet currents. Solid symbols: in-phase response; open symbols: quadrature response. The curves are fits to Eqtn.(1).

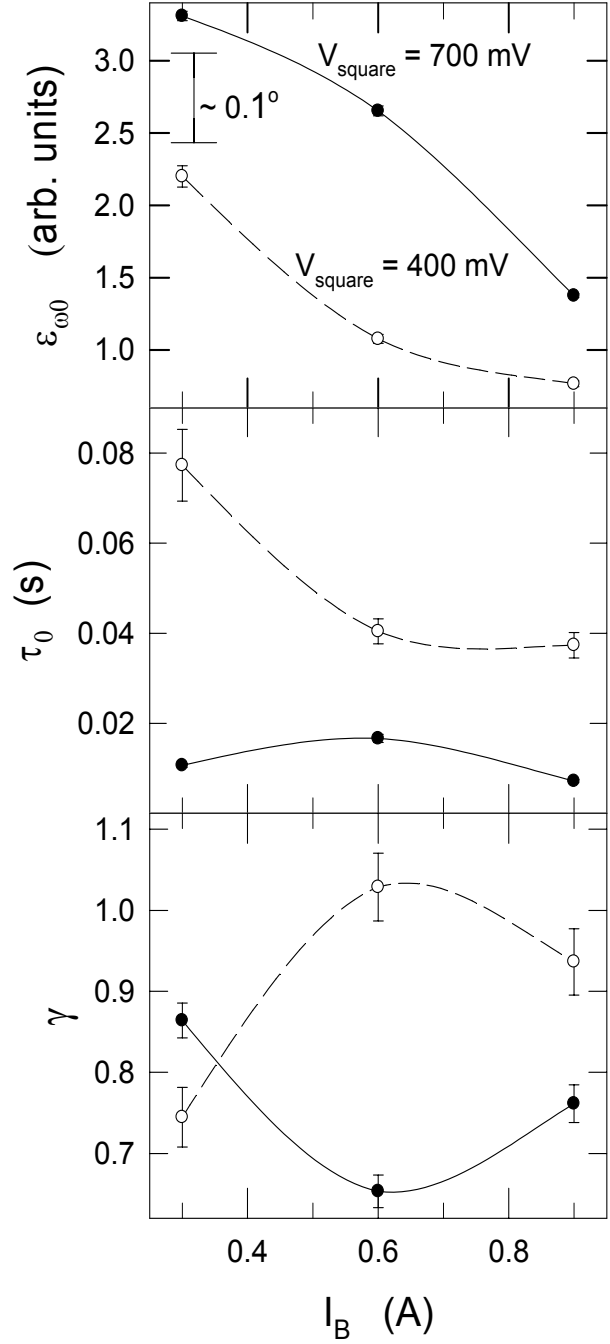

FIG. 3. Dependence of fitting parameters of Eqtn. (1) on magnet current for Sample E at T = 78 K (with twist angle $\partial\varphi/\partial I_B \sim 12°/\text{amp}$). Solid symbols: 700 mV square-waves; open symbols: 400 mV square-waves. The curves are guides to the eye. (Where not visible, error bars are smaller than the points.)

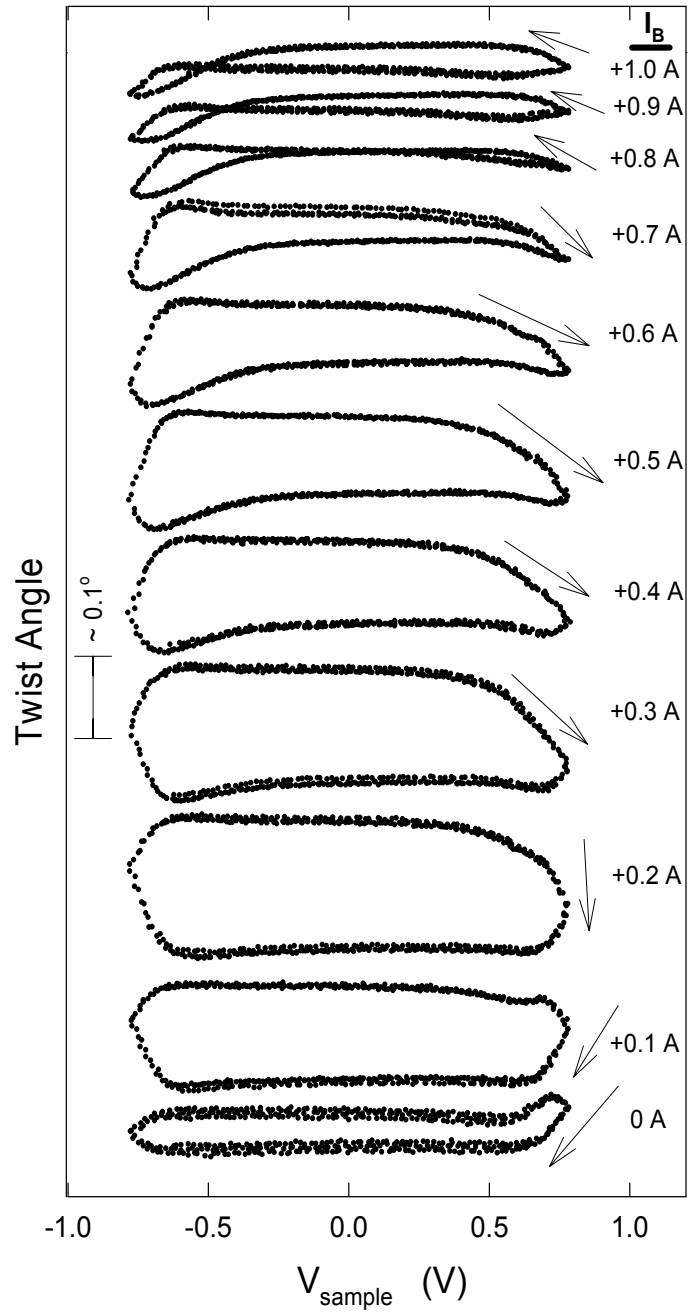

FIG. 4. VITS hysteresis loops measured for Sample F with 0.75 V, 0.3 Hz triangle waves at several magnet currents (with twist angle $\partial\varphi/\partial I_B \sim 5°$/amp) at T = 78 K. Curves for successive values of $I_B$ are offset for clarity. Arrows show the directions of the loops. (Three loops are overlaid for each value of $I_B$.)

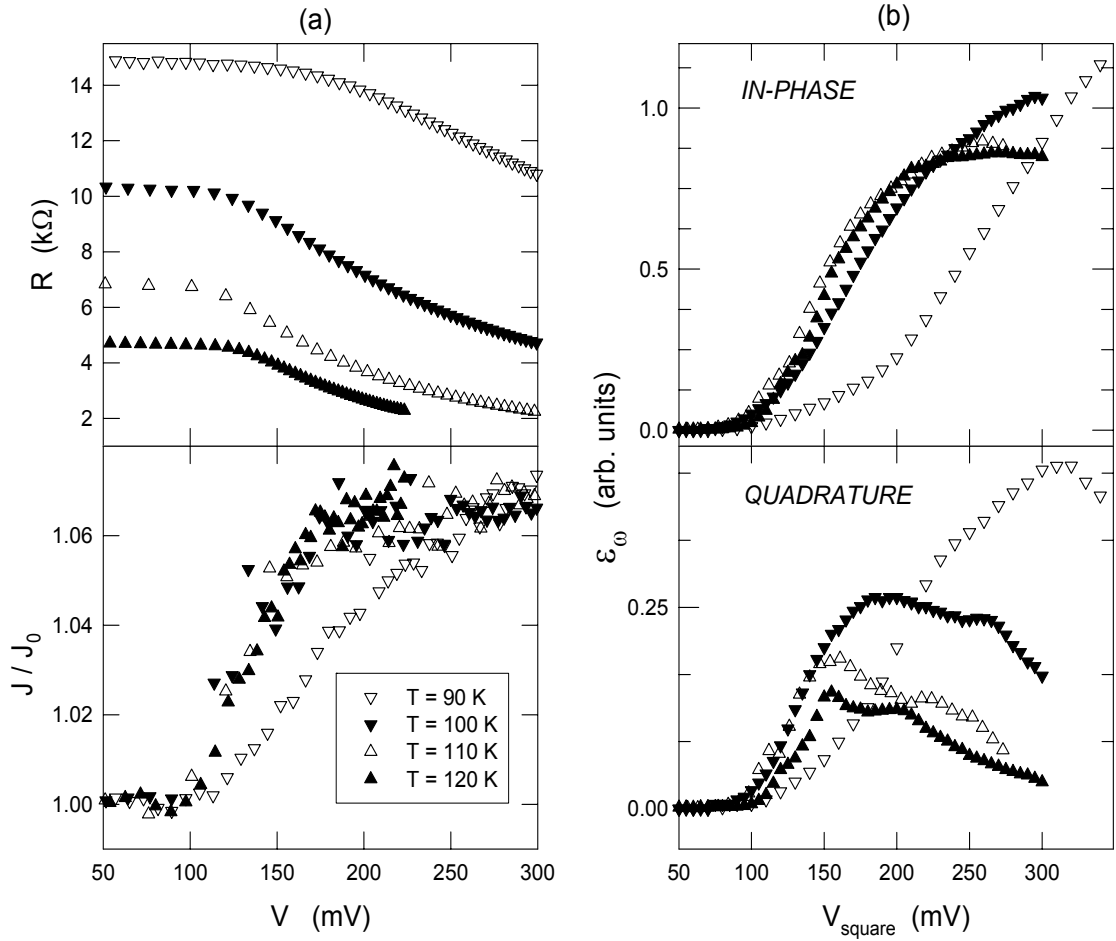

FIG. 5. (a) Resistance and shear compliance (measured with 10 Hz oscillating torque) vs. dc voltage at a few temperatures, measured for Sample G. (b) 10 Hz VITS response of Sample G vs. square-wave voltage at a few temperatures; the responses in-phase (top panel) and in quadrature (bottom panel) with the applied square-wave are shown.

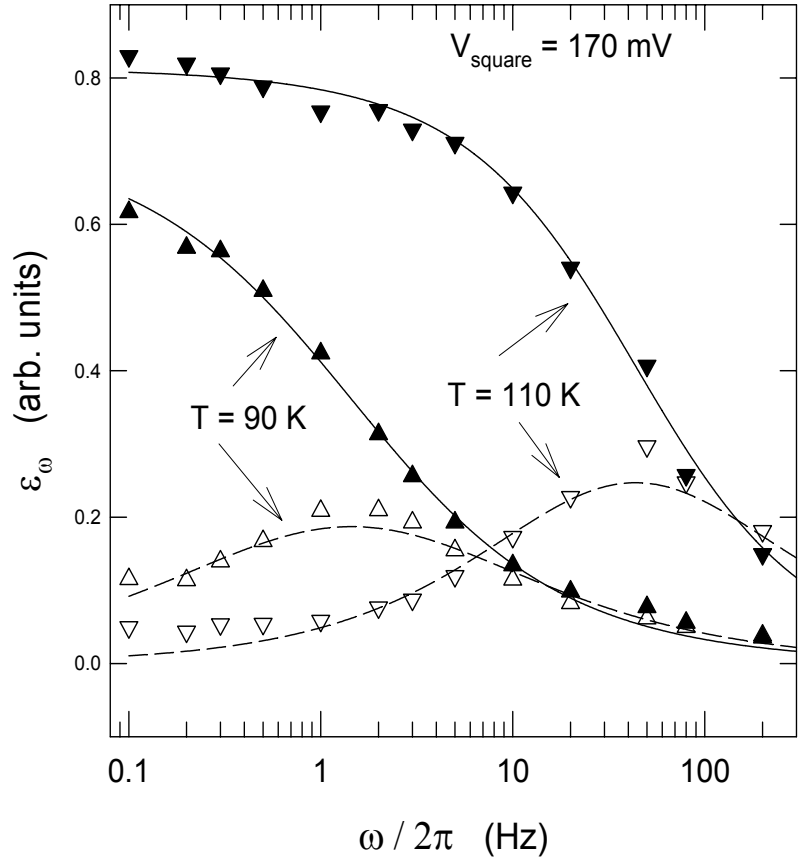

FIG. 6. Square-wave frequency dependence of $\varepsilon_\omega$ as function of frequency with $V_{square} = V_{on} + 100$ mV $= 170$ mV at two different temperatures for Sample G. Solid symbols: in-phase response; open symbols: quadrature response. The curves are fits to Eqtn.(1).

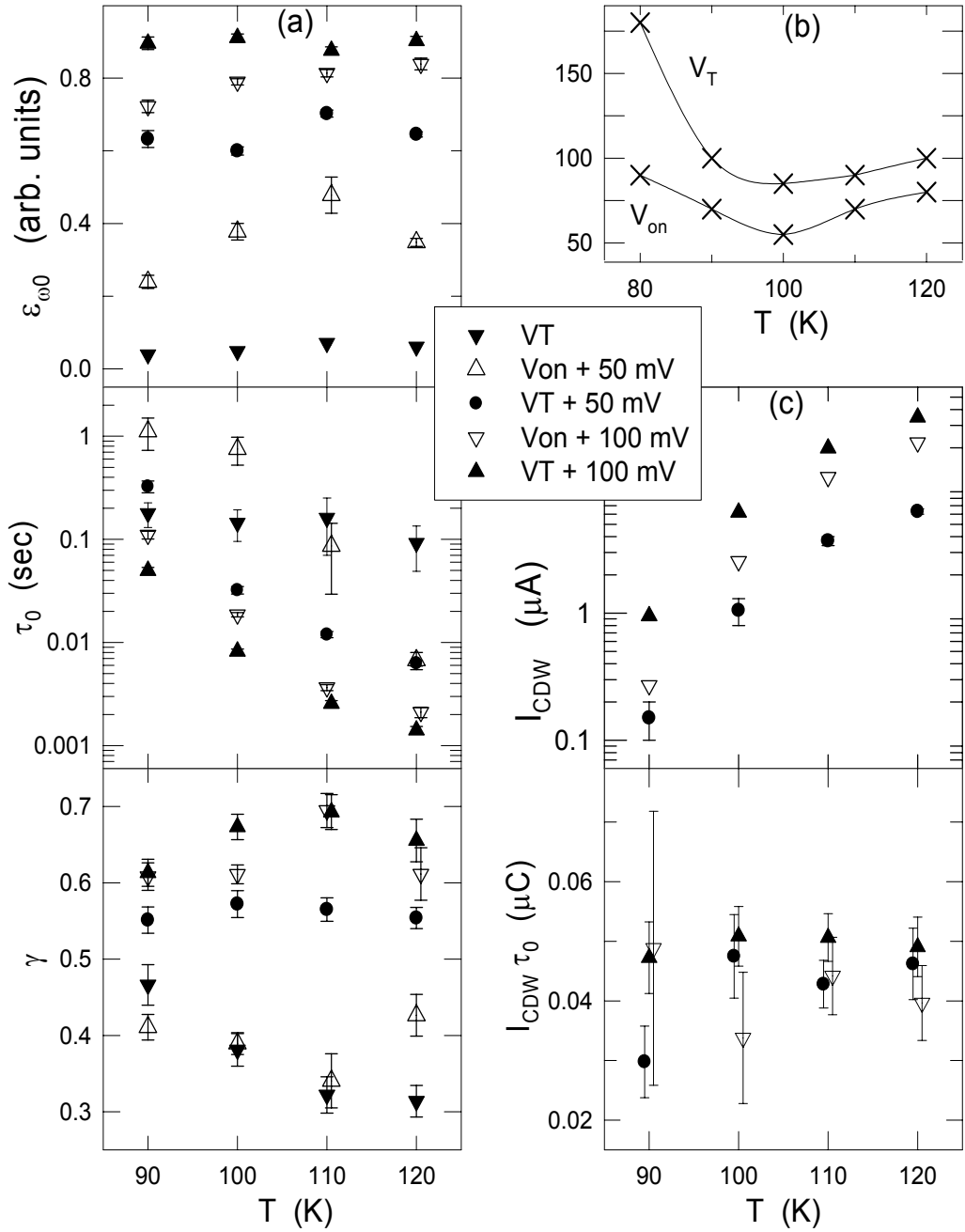

FIG. 7. (a) Fitting parameters to Eqtn. (1) for sample G vs. temperature for several voltages. (b) Threshold and onset voltages vs. temperature for Sample G; curves are guides to the eye. (c) CDW current and $I_{CDW}\tau_0$ vs. temperature at a few voltages. (When not shown, the error bars in $I_{CDW}$ are smaller than the symbols.)

22